\documentclass[11pt,preprint]{aastex}
\usepackage{times}

\newcommand{\vect}[1]{\ensuremath{\mbox{\boldmath $#1$}}}

\slugcomment
{submitted to the Astrophysical Journal (revised version)}
\shortauthors{AN \& HAN}
\shorttitle{WIDE BINARY LENS MICROLENSING}

\begin{document}

\title{Effect of a wide binary companion to the lens\\ 
on the astrometric behavior of gravitational microlensing events}

\author{Jin H. An}
\affil{Department of Astronomy, the Ohio State University, 
140 West 18th Avenue, Columbus, OH 43210}
\email{jinhan@astronomy.ohio-state.edu}
\and
\author{Cheongho Han}
\affil{Department of Physics, Chungbuk National University,
Chongju 361-763, Korea}
\email{cheongho@astroph.chungbuk.ac.kr}

\begin{abstract}
In this paper, we investigate the effect of a wide binary companion 
of the lens on the astrometric behavior of Galactic gravitational 
microlensing events and compare it to the effect on the photometric 
behavior.  We find that the wide binary companion of the lens can 
affect the centroid motion of images substantially even if the 
corresponding light curve appears to be the one of a standard single 
point-mass lens event.  The relatively significant effect of the wide 
binary lens on the astrometric lensing behavior, on one side, calls 
for careful consideration of the lens binarity in analyzing the future 
astrometric lensing data.  On the other side, larger astrometric effect 
of the companion makes astrometric lensing an efficient method to detect 
binary lenses over a broad range of separations.
\end{abstract}


\section{Introduction}
Light curves of microlensing events caused by a single point-mass lens
are characterized by non-repeating, single peak, and symmetric curves
\citep{paczynski86}, known as Paczy\'nski curves. On the other hand, if 
events are caused by a binary lens, the resulting light curves deviate
from those of single lens events.  The deviation caused by the binary 
nature of the lens becomes prominent when the angular separation between 
the two lens components (binary separation) is comparable to the angular 
Einstein ring radius of the lens system, which is related to the physical 
parameters of the system by
\begin{eqnarray}
\theta_{\rm E}
&=&
\left(\frac{2R_{\rm Sch}}{D}\right)^{1/2}
\nonumber\\
&\sim&0.72\ \mbox{mas}
\left(\frac{D_{\rm S}}{D_{\rm L}}-1\right)^{1/2}
\left(\frac{D_{\rm S}}{8\ \mbox{kpc}}\right)^{-1/2}
\left(\frac{M}{0.5\ \mbox{M$_\sun$}}\right)^{1/2}
\label{eq1}
\end{eqnarray}
where $D^{-1}=D_{\rm L}^{-1}-D_{\rm S}^{-1}$, $D_{\rm L}$ and $D_{\rm S}$ 
are the distances to the lens and the source from the observer respectively, 
and $R_{\rm Sch}$ is the Schwarzschild radius corresponding to the total 
mass $M$ of the lens system.  As the binary separation becomes larger, 
however, the individual lens components begin to act as if they are two 
independent single lenses.  For these wide binary lens systems, it is 
likely that the source trajectory will pass close only to one of the lens
components and the resulting light curve will look very similar to a standard
Paczy\'nski curve. If the Galactic binaries have a similar distribution of
binary separations to that of binaries in the solar neighborhood 
\citep{duquennoy91}, the majority of the binaries located along the lines of
sight towards the Galactic bulge will have separations substantially larger
than their Einstein ring radii whose typical values are several hundred
micro- to a few milli-arcseconds. Therefore, for most of Galactic binary
lens events, it will be difficult to identify the existence of the
companion to the lens from the analysis of the light curves alone except 
for the rare case of either caustic crossing or the source star's passage 
near both lens components \citep[a repeating event;][]{stefano99}.

As an alternative method of microlensing observations, it was proposed to
measure the lensing-induced displacement of the (brightness-weighted) 
image centroid position (usually referred to centroid shifts) by high 
precision astrometry \citep*{miyamoto95,hog95,walker95,escude96,paczynski98,
boden98,han99, dominik00}. When a source star is microlensed, it becomes 
split into several images whose number depends on the structure of the 
lens system, e.g., two images for a single lens and three or five images 
for a binary lens system.  The typical separation between the images is 
an order of $\theta_{\rm E}$, and thus it is impossible to resolve the 
individual images even with the highest angular resolution achieved so far.  
However, several planned high-precision astrometric satellite experiments,
e.g., {\it Space Interferometry Mission} (SIM), {\it Double Interferometer 
for Visual Astrometry} (DIVA), and {\it Global Astrometric Interferometer 
for Astrophysics} (GAIA), and optical/NIR interferometers on 10-m class 
ground-based telescopes, e.g., the Keck interferometer, the Very Large 
Telescope Interferometer (VLTI), and the Large Binocular Telescope (LBT), 
will have precisions in position measurements of the order of tens of 
$\mu$-arcsecs or better, and thus will enable one to measure the 
lensing-induced centroid shifts.

If an event is caused by a point-mass lens, the trajectory of the centroid
shifts traces an ellipse (see \S~\ref{sec2}), known as an astrometric ellipse
\citep*{walker95,jeong99,dominik00}.  If an event is caused by a binary 
lens, on the other hand, the centroid shift trajectory deviates from the 
elliptical one of a single lens event, which is characterized by
distortions, twistings, and jumps \citep*{han99_2}.

In this paper, we investigate how a wide binary companion of the lens
affects the observed photometric and astrometric microlensing behaviors.  
From this investigation, we find that the effect on the trajectory of 
the image centroid shifts is significantly more important than on the 
light curve.  Hence, we argue that a careful consideration of the effect 
of a wide binary lens will be required in analyzing the measured centroid
shifts even if the corresponding light curve appears to be a standard
Paczy\'nski curve.

\section{Basics of Microlensing
\label{sec2}}

If a source located at $\vect{r}_{\rm S}$ (on the projected plane of the 
sky) is lensed by a $N$-point-mass lens system, where the individual 
components' masses and locations are $m_i$ and $\vect{r}_{{\rm L},i}$, 
the positions of the images $\vect{r}$ are obtained by solving the equation 
of lens mapping (the lens equation);
\begin{equation}
\label{eq2}
\vect{r}_{\rm S}=\vect{r}-\theta_{\rm E}^2
\sum_{i=1}^N\frac{m_i}{M}
\frac{ \vect{r}-\vect{r}_{{\rm L},i}}
{|\vect{r}-\vect{r}_{{\rm L},i}|^2}
\,,\end{equation}
where $M=\sum_i^N m_i$.  Since the lensing process conserves the surface
brightness (i.e.\ does not create nor destroy photons), the flux ratio
between the lensed image and the unlensed source is simply given by the
surface area ratio between the image and the unmagnified source, i.e., 
the magnification.  For a point source limit, the magnifications of the 
individual images are given by the inverse of the Jacobian determinant of 
the lens mapping evaluated at each image position $\vect{r}_j$;
\begin{equation}
A_j = 
\left(\frac{1}{|\det J|}\right)_{\vect{r}=\vect{r}_j}
\ ; \ \ \
\det J=\left|\frac{\partial\vect{r}_{\rm S}}{\partial\vect{r}}\right|
\,\end{equation}
and the total magnification is given by the sum of the magnifications
of the individual images, i.e., $A_{\rm tot}=\sum_j^{N_{\rm I}} A_j$,
where $N_{\rm I}$ is the total number of images. The location of the 
apparent image centroid corresponds to the magnification weighted mean 
of the individual image positions, and therefore, the centroid shift is 
given by
\begin{equation}
\vect{\delta}= 
\sum_j^{N_{\rm I}}\frac{A_j}{A_{\rm tot}}\vect{r}_j-\vect{r}_{\rm S}
\ .\end{equation}

For a single point-mass lens ($N=1$), the lens equation (eq.~[\ref{eq2}]) 
is easily solvable and solving the equation yields two image positions 
($N_{\rm I}=2$).  The total magnification and the centroid shift of the 
single lens event are expressed in analytical forms of
\begin{equation}
\label{eq5}
A=\frac{u^2+2}{u\sqrt{u^2+4}}
\,;\end{equation}
\begin{equation}
\label{eq6}
\vect{\delta}=\frac{\theta_{\rm E}\vect{u}}{u^2+2}
\,,\end{equation}
where $\vect{u}\equiv(\vect{r}_{\rm S}-\vect{r}_{\rm L})/\theta_{\rm E}$
is the dimensionless lens-source separation vector normalized by 
$\theta_{\rm E}$. Since the source and the lens move relative to each 
other, $\vect{u}$ changes with time. Under the approximation that this 
relative motion is rectilinear, it is represented by
\begin{equation}
\vect{u}= 
\left(\frac{t-t_0}{t_{\rm E}}\right)\hat{\vect{x}}+\beta\hat{\vect{y}}
\,,\end{equation}
where $t_{\rm E}$ represents the time required for the source to transit
$\theta_{\rm E}$ (Einstein time scale), $\beta$ is the closest lens-source
separation in units of $\theta_{\rm E}$ (impact parameter), $t_0$ is the
time at that moment, and the unit vectors $\hat{\vect{x}}$ and $\hat{\vect{y}}$
are parallel and perpendicular to the direction of the relative 
lens-source motion, respectively. The trajectory of the centroid shifts is 
an ellipse with a major axis and an eccentricity of 
$a=(\theta_{\rm E}/2)\times (\beta^2+2)^{-1/2}$
and $e=[(\beta^2/2)+1]^{-1/2}$ and its semi-major axis lies parallel to 
$\hat{\vect{x}}$.

For a multi-lens system ($N\ge 2$), on the other hand, the lens equation
cannot be solved algebraically. Hence, to obtain the image positions, one 
has to solve the equation numerically.  In particular, for the case of a 
binary lens ($N=2$), the lens equation can be expressed as a fifth-order 
complex polynomial \citep{witt90} and there exist three or five images 
depending on the source position with respect to the lens positions.

\section{Effects of a Wide Binary Lens Companion}

The existence of a binary companion to the lens affects both the photometric
and astrometric behaviors of a microlensing event.  The most important 
effect of a wide binary companion of the lens is that it makes the 
effective position of the primary\footnote{Here, the primary denotes 
the binary component which lies closer to the projected path of the source 
on the sky, and the use of `the companion' is reserved for the other 
component of the binary.  We note, however, that the basic results of 
the subsequent discussion do not change under the usual definition of 
the primary as the more massive binary component.} move towards it or, 
equivalently, the effective source position move with respect to the 
primary \citep{stefano96}.  From the lowest order approximation of 
equation~(\ref{eq2}), one finds that this effect is well characterized by 
the use of an {\it effective lens-source separation vector}, which is 
expressed by
\begin{equation}
\label{eq8}
\tilde{\vect{u}}\equiv
\vect{u}_1
-\frac{q}{d}
\frac{\vect{r}_{{\rm L},2}-\vect{r}_{{\rm L},1}}
{|\vect{r}_{{\rm L},2}-\vect{r}_{{\rm L},1}|}
\,,\end{equation}
where the subscripts `1' and `2' represent the primary and the companion
respectively, $q = m_2/m_1$ is the mass ratio, and 
$\vect{u}_1 = (\vect{r}_{\rm S}-\vect{r}_{{\rm L},1})/\theta_{{\rm E},1}$,
and $d = |\vect{r}_{{\rm L},2}-\vect{r}_{{\rm L},1}|/\theta_{{\rm E},1}$
are the distances of the source and the companion from the primary lens 
normalized by the Einstein ring radius of the primary $\theta_{{\rm E},1}$, 
respectively. We note that this effect is not generally measurable because
it requires an a priori information on the actual positions of the lens
components on the sky.

As an additional effect of the binary lens companion, one may expect that 
the observed light curve and the centroid shift trajectory would be the 
superposition of those resulting from the two events in which the component 
lenses behave as two independent lenses. However, this superposition 
effect of the lens companion is important only for the trajectory of the 
centroid shifts but not for the light curve of an event.  This can be seen 
from the magnification and the centroid shift caused by the companion when 
the source is located close to the primary (i.e., $\vect{r}_{\rm S}\simeq
\vect{r}_{{\rm L},1}$), which are, respectively, represented by
\begin{equation}
\label{eq9}
A_2\sim 1+2\frac{q^2}{d^4} \ \ \ ({\rm for}\ d\gg 2)
\,,\end{equation}
\begin{equation}
\label{eq10}
\delta_2\sim
\theta_{{\rm E},1}\frac{q}{d} \ \ \ ({\rm for}\ d\gg \sqrt{2})
\ .\end{equation}
From equations~(\ref{eq9}), one finds that as the binary separation 
becomes larger, the photometric superposition effect falls off rapidly
($\sim d^{-4}$): faster than the decrease of $|\tilde{\vect{u}}-\vect{u}_1|$ 
($\sim d^{-1}$; see eq.~[\ref{eq8}]). On the other hand, the astrometric 
effect endures up to a very large separation at which the photometric 
effect is negligible (the same order effect as the positional shift; 
see eq.~[\ref{eq10}]).

\begin{figure}
\plotone{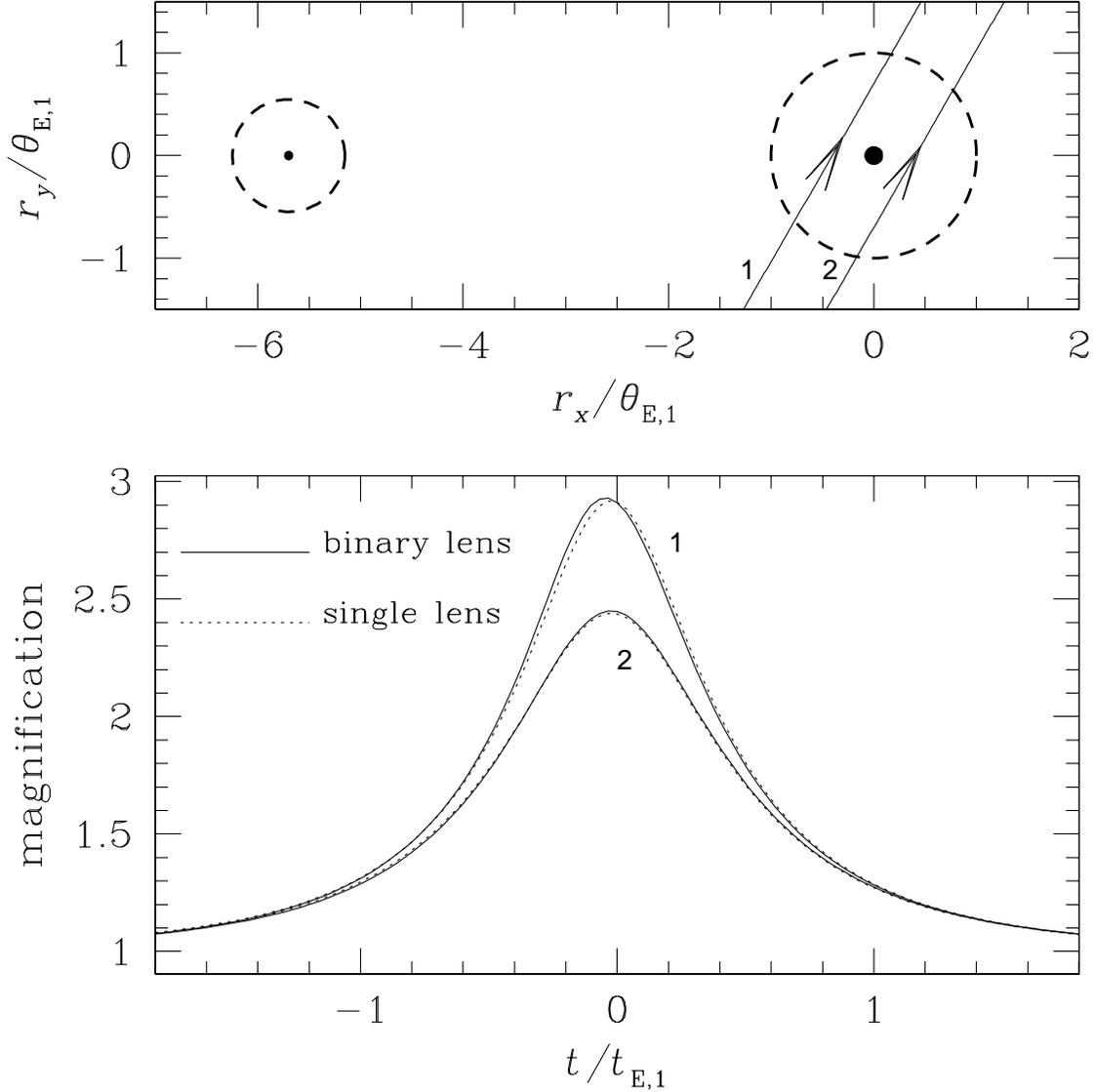}
\caption{
Example light curves of microlensing events caused by a wide binary lens.
The upper panel shows the geometry of the lens system. The small filled dots
represent the primary (right) and the companion (left) and the circles drawn
by dashed lines are the Einstein rings associated with the individual lenses.
All coordinate scales are normalized to the Einstein ring radius of the
primary $\theta_{{\rm E},1}$. The straight lines with arrows represent the
source trajectories projected onto the sky.  In the lower panel, we present
light curves ({\it solid} curves) resulting from the two source trajectories
marked by numbers in the upper panel.  Also drawn are the light curves (dotted
curves) of the events where the primary acts as a single lens at the effective
position accounting the presence of the companion to the lens
(see eq.~[\ref{eq8}]).
\label{fig1}}\end{figure}

\begin{figure}
\plotone{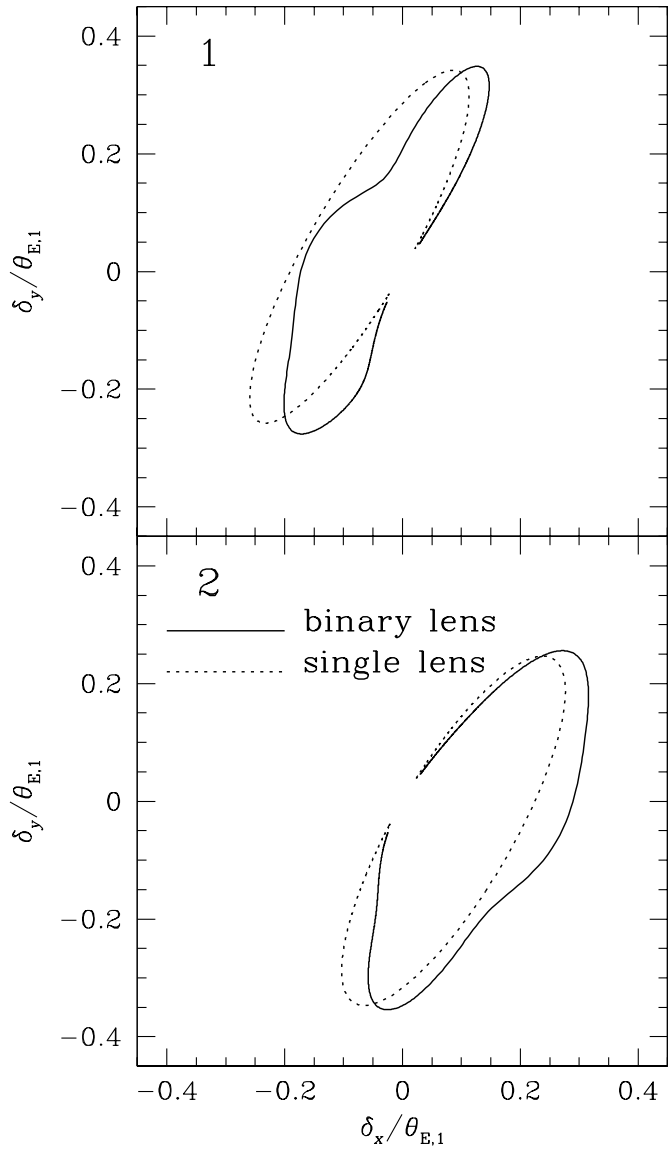}
\caption{
Trajectories of the image centroid shifts for the same events whose light
curves are presented in Fig.~\ref{fig1}.  The line types are selected so 
that they match with those of the corresponding light curves.  The number 
in each panel corresponds to that used to designate the source trajectory 
in the upper panel of Fig.~\ref{fig1}.
\label{fig2}}\end{figure}

To illustrate the larger astrometric superposition effect of the lens 
companion than its photometric effect, we present the light curves 
(Fig.~\ref{fig1}) and the trajectories of the image centroid shifts 
(Fig.~\ref{fig2}) of events caused by a wide binary lens with the mass 
ratio $q=0.3$ and the separation $d=5.7$. One finds that the light curves 
of these events are well approximated by that of a single lens event after 
accounting for the effective positional shift of the primary.  By contrast, 
the corresponding centroid shift trajectories deviate significantly from 
an elliptical one expected for a single lens event.

\section{Detection of Wide Binary Lens Companions}

In order to assess how well one can distinguish light curves of wide binary
lensing events from the ones of single point-mass events, we define 
an magnification excess by
\begin{equation}
\label{eq11}
\epsilon=\frac{A-\tilde{A}_0}{\tilde{A}_0}
\,,\end{equation}
where $A$ is the exact magnification of the wide binary lens event and 
$\tilde{A}_0$ is that of the single lens event with a mass equal to the 
primary and located at the effective position defined by equation~(\ref{eq8}),
\begin{equation}
\tilde{A}_0=\frac{\tilde{u}^2+2}{\tilde{u}\sqrt{\tilde{u}^2+4}}
\ .\end{equation}

\begin{figure}
\plotone{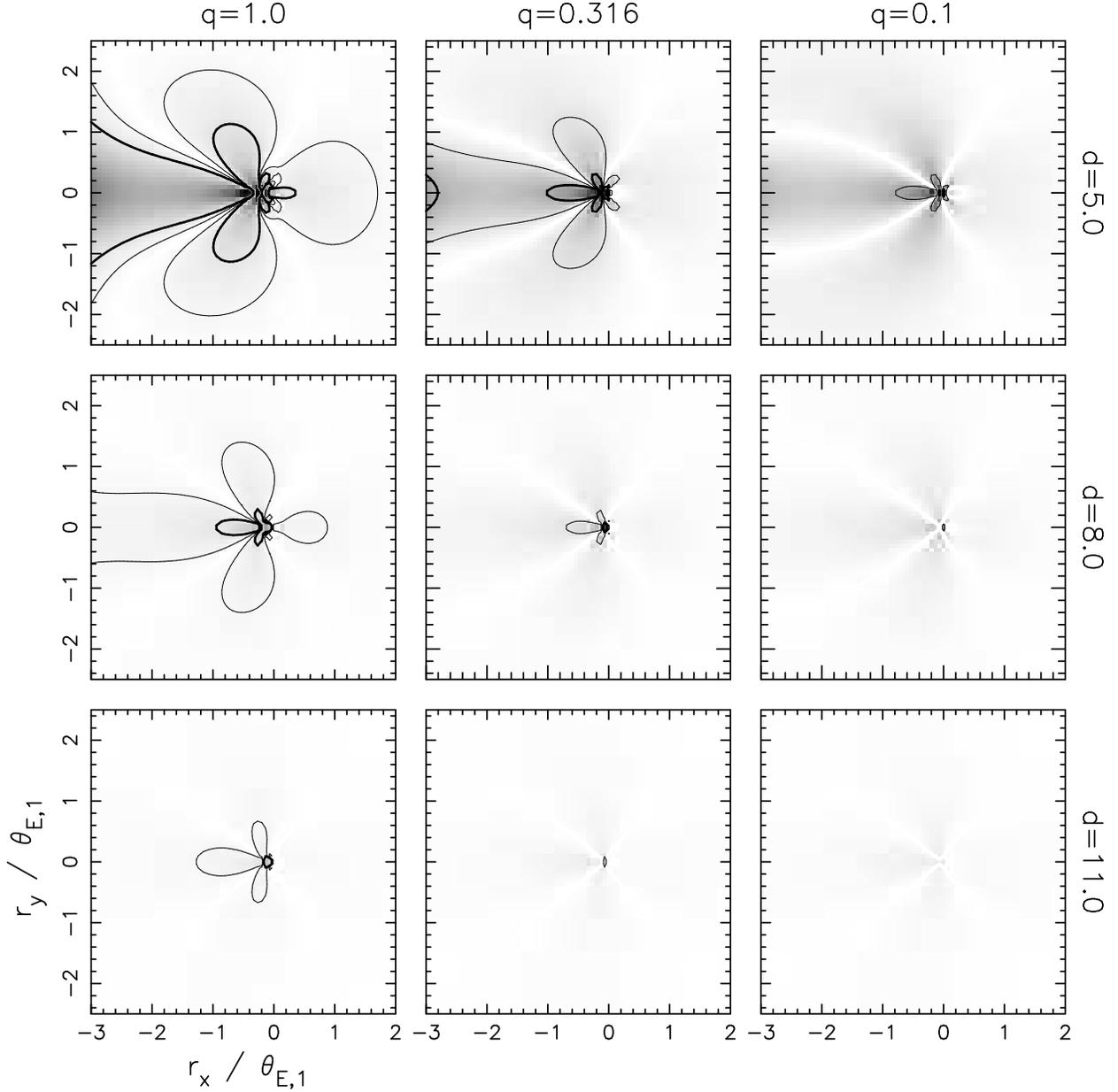}
\caption{
Contour maps of magnifications excess, $\epsilon$ (see eq.~[\ref{eq11}] 
for the definition).  The binary separations and the mass ratios of the 
binary lenses shown are $q=1.0,\ 0.316,\ 0.1$ and $d=5.0,\ 8.0,\ 11.0$. 
All scales are normalized by the Einstein ring radius of the primary, 
$\theta_{{\rm E},1}$.  The origin is the position of the primary and the 
companion is located on the left side of the primary, i.e.\ ($-d$, 0).  
Contours are drawn at the levels of $\epsilon=\pm$3\% ({\it thick lines}) 
and $\pm$1\% ({\it thin lines}).  In addition to the contours, grey scale 
is used to provide the idea of the gradient over the region.
\label{fig3}}\end{figure}

Figure~\ref{fig3} shows the magnification excess map of the region close 
to the primary lens for wide binary lens systems with various values of 
$d$ and $q$. From the figure, one finds that the region of significant 
deviations shrinks rapidly as $d$ increases.  Noticeable deviations 
(e.g., $\epsilon\ga 2$\%) occur only if the source passes very close to 
the primary, implying that the presence of the companion to the lens can 
be detected effectively only if the source crosses the caustic associated 
with the primary or passes the region of the cusp-influence magnification. 
However, both the size of the caustic and the strength of the cusp converge 
to those of a point-mass lens much faster than the effect of the positional 
shift.  As a result, even with the photometric precision as good as 0.01 
mag and sufficiently dense sampling during the peak magnification, the 
binary detection efficiency is expected to be low for most wide binary 
microlensing events with $d\ga 10$.

In order to compare the photometric deviation to the one expected from
astrometric observations, we also define the deviation of centroid shift by
\begin{equation}
\label{eq13}
\Delta\vect{\delta}=\vect{\delta}-\tilde{\vect{\delta}}_0
\,,\end{equation}
where $\vect{\delta}$ is the centroid shift of the wide binary lensing 
event while $\tilde{\vect{\delta}}_0$ represents the centroid shift 
caused by the primary alone placed at its effective position,
\begin{equation}
\tilde{\vect{\delta}}_0=
\frac{\theta_{{\rm E},1}\tilde{\vect{u}}}{\tilde{u}^2+2}
\ .\end{equation}

Figure 4 shows the contour maps of $\Delta\delta$ of the same region and 
for the same binaries as in Figure 3.  When the source is located near 
the primary, the centroid shift deviation -- mostly due to the superposition 
effect -- is an order of $\sim (q/d)\theta_{\rm E,1}$ (see eq. [10]). 
Therefore, we draw contours at the levels which are scaled by 
$(q/d)\theta_{\rm E,1}$.  Figure 5 show the maps in a wider region where 
the contours are drawn at 6 different levels ranging $\sim 1\%$ to
$\sim 25\%$ of $\theta_{{\rm E},1}$.
Considering that the size of the angular Einstein ring radius is an order 
of a few milli- to several hundred micro-arcseconds for events caused by 
Galactic G, K, and M dwarfs with masses in the range of 
$\sim 0.1\ \mbox{M$_\sun$}$ -- $1\ \mbox{M$_\sun$}$ (see eq.~[\ref{eq1}]), 
the deviations for the events caused by a binary lens even with 
$q/d \sim 0.01$ are expected to be still several micro-arcseconds, which 
is comparable to the expected astrometric precision of the new generation 
interferometers.\footnote{The astrometric precision of SIM, for instance, 
will be as good as $5\ \mbox{$\mu$as}$ for stars brighter than 
$V=20$ \citep*{unwin97}.} Moreover, large deviations are found throughout 
the map, rather than concentrating near the primary as in the map of 
magnification excess.  Notably, quite complex patterns of deviation are 
present over the region of the circle around the primary with the radius 
of about $\theta_{\rm E,1}$ even for the case of $d\sim 10$.  This implies 
that the astrometric deviation will typically last for $\sim t_{\rm E}$.  
From astrometric microlensing observations of events, therefore, one can 
detect the presence of binary companions to the lens for a broader range 
of separations with an increased efficiency than from conventional 
photometric observations, possibly even with less dense sampling.

\begin{figure}
\plotone{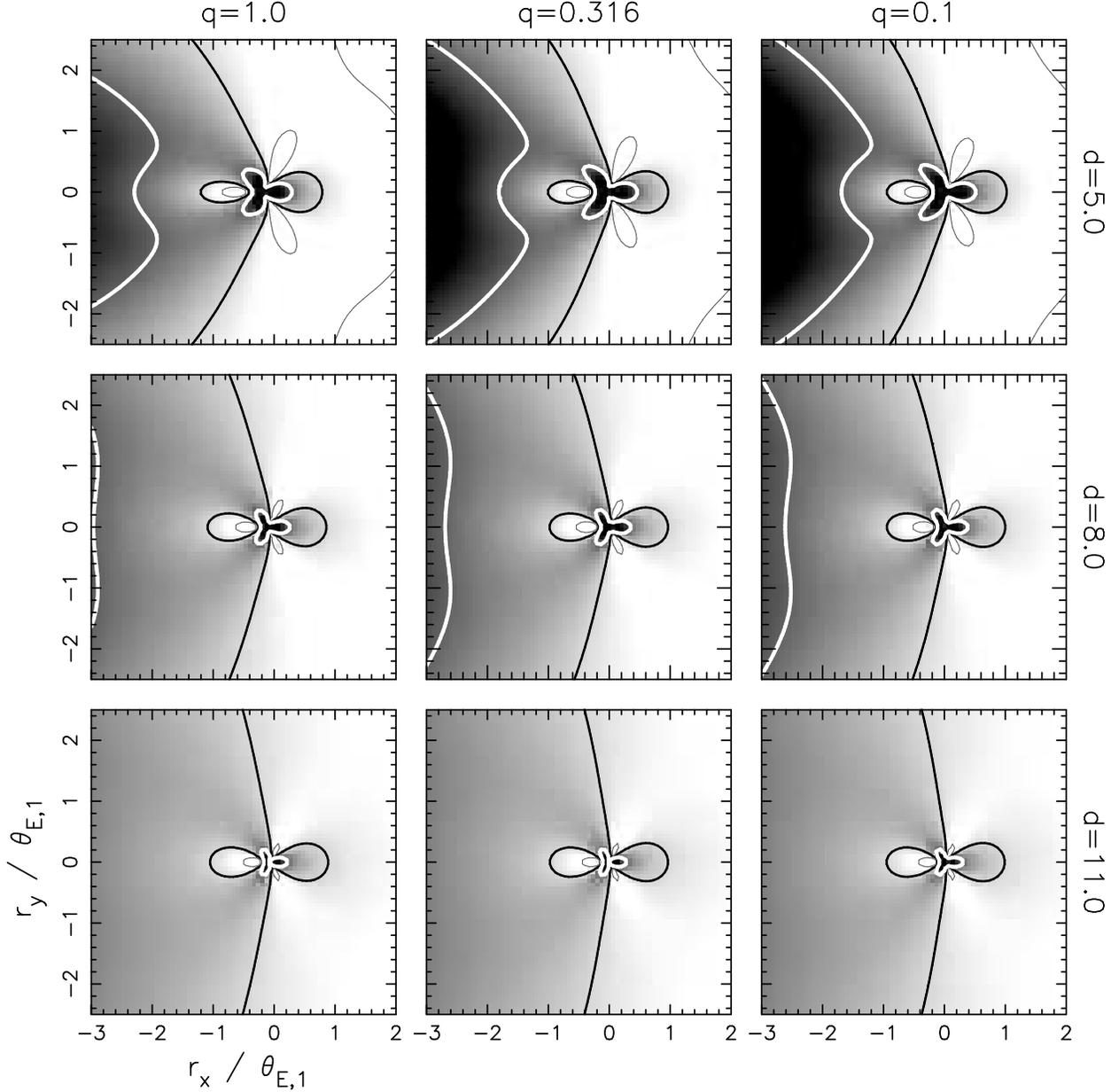}
\caption{
Contours of centroid shift deviations, $\Delta\delta$ (see eq.~[\ref{eq13}]),
in the same region and for the same wide binaries as in Fig.~\ref{fig3}.
The coordinate systems are also the same as in Fig.~\ref{fig3}.
Contours are drawn at the levels of
$\Delta\delta = (2/3)\times(q/d)\times\theta_{{\rm E},1}$
({\it thin black lines}),
$q/d\times\theta_{{\rm E},1}$ ({\it thick black lines}),
and $(3/2)\times(q/d)\times\theta_{{\rm E},1}$ ({\it thicker
white lines}).  Grey scale is used to provide the idea of the gradient
over the region, but does not necessarily correspond to the absolute values
of deviations.
\label{fig4}}\end{figure}

As a quantitative comparison between the detection efficiencies of wide 
binary companions to the lens expected from the photometric and the 
astrometric lensing observations, we perform a Monte-Carlo simulation.  
In the simulation, we produce a large number of wide binary lens events
and fit their light curves and image centroid shift trajectories to those 
of single point-mass lens events (eqs.~[\ref{eq5}] and [\ref{eq6}]).
We assume the lens binarity is detected if the resulting $\chi^2$ of the 
fit is larger than a statistically acceptable value (see below).  
Tables~\ref{tab1} and \ref{tab2} list the photometric and astrometric
detection rate determined by the simulation, each for the same wide 
binaries as Figures~\ref{fig3} and \ref{fig4} under two different sampling 
strategies and three different measurement errors.  The resulting detection 
rate in each entry is based on 5,000 trial simulations.  We note that, 
since we are more interested in the possibility of the inference to the 
presence of the lens companion, rather than in the determination of the 
absolute detection efficiency, we exclude the repeating events and the 
caustic-crossing events for which the binarity of the lens can be 
immediately inferred.  Trial events are generated with random trajectories 
for each fixed configuration of the binary (i.e, at fixed $d$ and $q$),
but restricted only to high-magnification,\footnote{The maximum 
magnification due to the primary is greater than two, i.e.\ 
$\min\{|\vect{r}_{\rm S}-\vect{r}_{{\rm L},1}|^2)\} <[(4/\sqrt{3})-2]
\theta_{\rm E,1}^2$.} non-repeating,\footnote{The magnification due to 
the companion is not bigger than two, i.e.\ $\min\{|\vect{r}_{\rm S}-
\vect{r}_{{\rm L},2}- (\vect{r}_{{\rm L},1}-\vect{r}_{{\rm L},2})/d^2|^2\}
\ge[(4/\sqrt{3})-2]\theta_{{\rm E},2}^2$.} and non-caustic-crossing 
events.\footnote{The number of data points when the source is inside the 
caustic is, at most, one.} To simulate light curves, we fix the unlensed 
source flux at an arbitrary value and include also random amounts of 
blended light of less than 20\% of the unlensed source flux.  As for the 
observation, we examine two different types of sampling.  In the first 
type, we assume 200 photometric and astrometric measurements during 
$-t_{\rm E}\leq t_{\rm obs}-t_0 \leq 3 t_{\rm E}$ (`Broad' sampling).
In the second type, the same number of measurements are performed during 
a shorter period of time around the photometric peak of $-0.5 t_{\rm E} 
\leq t_{\rm obs}-t_0 \leq 1.5 t_{\rm E}$ (`Dense' sampling).  For 
simplicity, the sampling rate is taken to be uniform and the uncertainty 
associated with each measurement is the same throughout the measurements 
(i.e., the deviations distribute as Gaussian with a single variance). 
Finally, the best-fitting single lens model and $\chi^2$ of a simulated 
event are determined through the $\chi^2$ minimization by successive 
linearized fitting.  For light curves, we fit five parameters: $t_0$, 
$t_{\rm E}$, $\beta$, and the unlensed source flux and the blended light 
while, for the image motion, we fit seven parameters: $t_0$, $t_{\rm E}$, 
$\beta$, $\theta_{\rm E}$, and the direction of the relative lens-source 
motion and the lens position on the sky (which is characterized by two 
parameters).  The detection is defined by
\begin{equation}
\label{eq15}
\sqrt{2\chi^2}-\sqrt{(2\cdot{\rm dof}-1)}\ge 3
\,,\end{equation}
where `${\rm dof}$' denotes the degrees of freedom of the fitting.
The criterion was chosen to be the 3$\sigma$ deviation of $\chi^2$ from
the expectation value if the underlying model (i.e., single point-mass
lens event) were correct. However, due to the positive skewness of the
$\chi^2$-distribution, the criterion actually corresponds to the rejection 
of the single lens model with the confidence level of approximately 
99.8\%.

As expected, the result shows a higher binary detection rate for the
10-$\mu$as astrometry ($\theta_{\rm E}\ga 0.5\ \mbox{mas}$) than for
the 1\% photometry.  For example, the astrometric detection efficiency
of a equal-mass binary with a separation of $d\sim 10$ stays more than
50\% even for events with $\theta_{\rm E}\sim 0.5\ \mbox{mas}$, while
the photometric detection rate falls below 10\% (2\% photometry) to 30\%
(0.5\% photometry).  Furthermore, even in the regime where the photometric
deviations are basically indistinguishable from random errors, the
astrometric  deviations can be unambiguously detected for a substantial
fraction (5$\sim$10\%) of events under a reasonable accuracy and a proper
sampling strategy.

Another interesting result one finds from Tables~\ref{tab1} and \ref{tab2}
is that `Broad' sampling strategy has an advantage in detecting wide binary
lens companions over `Dense' sampling strategy if the binary effect is strong
(i.e., smaller $d$ and/or larger $q$) and/or the measurement precision
is high.  On the other hand, `Dense' sampling strategy becomes advantageous
over `Broad' sampling strategy as the binary signal becomes weaker (i.e.,
larger $d$ and/or smaller $q$) and/or the measurement precision is poor.
It is also interesting to see that the relative advantage of `Dense' sampling
strategy for weak binaries and/or poor measurements is more obvious for the
photometry than the astrometry.  This is more or less expected by the
examination of the deviation maps in Figures~\ref{fig3} and \ref{fig4},
i.e., the region of substantial deviation is much more wide-spread in the 
astrometric deviation map than in the photometric one, and thus the 
photometric deviations will be likely to be shorter-lived than the 
astrometric deviations.

\section{Discussion}
The relatively significant effect of the wide binary lens companion on the
astrometric lensing behavior compared to the effect on the photometric
behavior, on one side, calls for careful considerations of the lens
binarity when one attempts to analyze and interpret the future astrometric
lensing data. In the current lensing experiments based on photometric 
observations, undetected wide binary lens companions do not pose as a 
serious problem because the resulting light curve is, in most cases, well 
approximated by that of a single lens event and more importantly the 
resulting time scale of the event (the only physically interesting lensing 
parameter for a single lens event) is hardly affected by the existence of 
the wide binary lens companion.  However, as demonstrated in the previous 
sections, one generally cannot ignore the effect of the wide binary lens 
companion on the observed centroid shift trajectory even if the 
corresponding light curve has little signature of the lens binarity.

The large astrometric effect of the wide binary lens companion, on the 
other side, makes the future astrometric lensing experiment a useful 
method to detect binary lens companions over a broader range of separations.  
Especially, in searches for sub-stellar-mass and/or sub-luminous companions,
astrometric lensing can provide a much more effective channel to detect 
them than the lensing light curve analysis.  In addition, the study of the 
binary frequency can benefit from the astrometric lensing observations since 
it is sensitive to the existence of the companion to the lens over a wider 
range of separations and consequently suffers less severe bias towards 
resonant separation than the conventional lensing technique.

One possible complication involving the interpretation of binary-affected
centroid shift trajectories is the parallax effect.  The parallax effect
becomes important due to the long-range duration of the centroid shifts
up to a large lens-source separation.  Therefore, proper interpretation of
the observed centroid motion of a wide binary lens event will require
careful consideration of the parallax effect.  We defer systematic studies
of the parallax effect for a future work.

\section{Conclusion}

By investigating the effect of wide binary lens companions on the photometric
and astrometric behaviors of microlensing events, we find that the signature
of the companion to the lens is significantly more important in the centroid 
shifts than in the light curve. On one side, this implies that, in analyzing 
the centroid shifts to be measured from the prospective astrometric lensing
experiments, one should consider the effect of wide binary lens companions
with more caution, which is generally not required in the analysis of light
curves obtained from the current-type photometric lensing experiments.
On the other side, this signifies the importance of the future astrometric 
lensing observations in providing a useful channel to search for binary 
companions to the lens over a broader range of separations.

\acknowledgements
This work was supported by a grant R01-1999-00023
from the Korea Science \& Engineering Foundation (KOSEF).
Work by J. An is supported by the Presidential
Fellowship from the Graduate School of the Ohio State University.

\clearpage

\begin{deluxetable}{ccrrr}
\tablecaption{
Photometric wide binary detection rate
\label{tab1}}
\tablehead{
\colhead{$d$}&
\colhead{photometric uncertainty (mag)}&
\colhead{$q=1.0$}&
\colhead{$q=0.316$}&
\colhead{$q=0.1$}}
\startdata
\cutinhead{`Broad' sampling\tablenotemark{a}}
  $5.0$&$5\times 10^{-3}$&$(87.9\pm 0.5)$\%&$(72.9\pm 0.6)$\%&$(30.6\pm 0.7)$\%\\
\nodata&      $0.01$     &$(74.7\pm 0.6)$\%&$(49.1\pm 0.7)$\%&$(11.3\pm 0.4)$\%\\
\nodata&      $0.02$     &$(60.3\pm 0.7)$\%&$(25.6\pm 0.6)$\%&$( 3.3\pm 0.3)$\%\\
  $8.0$&$5\times 10^{-3}$&$(63.9\pm 0.7)$\%&$(23.2\pm 0.6)$\%&$( 3.1\pm 0.2)$\%\\
\nodata&      $0.01$     &$(42.8\pm 0.7)$\%&$(10.1\pm 0.4)$\%&$( 0.6\pm 0.1)$\%\\
\nodata&      $0.02$     &$(23.0\pm 0.6)$\%&$( 4.3\pm 0.3)$\%&$( 0.2\pm 0.1)$\%\\
 $11.0$&$5\times 10^{-3}$&$(32.8\pm 0.7)$\%&$( 6.3\pm 0.3)$\%&$( 0.2\pm 0.1)$\%\\
\nodata&      $0.01$     &$(18.2\pm 0.5)$\%&$( 1.5\pm 0.2)$\%&$( 0.2\pm 0.1)$\%\\
\nodata&      $0.02$     &$( 9.8\pm 0.4)$\%&$( 0.5\pm 0.1)$\%&$( 0.2\pm 0.1)$\%\\
\cutinhead{`Dense' sampling\tablenotemark{b}}
  $5.0$&$5\times 10^{-3}$&$(78.9\pm 0.6)$\%&$(65.1\pm 0.7)$\%&$(32.3\pm 0.7)$\%\\
\nodata&      $0.01$     &$(68.5\pm 0.7)$\%&$(46.7\pm 0.7)$\%&$(15.9\pm 0.5)$\%\\
\nodata&      $0.02$     &$(62.3\pm 0.7)$\%&$(29.2\pm 0.6)$\%&$( 8.2\pm 0.4)$\%\\
  $8.0$&$5\times 10^{-3}$&$(59.3\pm 0.7)$\%&$(27.0\pm 0.6)$\%&$( 6.9\pm 0.4)$\%\\
\nodata&      $0.01$     &$(44.8\pm 0.7)$\%&$(13.9\pm 0.5)$\%&$( 2.1\pm 0.2)$\%\\
\nodata&      $0.02$     &$(32.0\pm 0.7)$\%&$( 7.3\pm 0.4)$\%&$( 0.9\pm 0.1)$\%\\
 $11.0$&$5\times 10^{-3}$&$(34.4\pm 0.7)$\%&$( 9.6\pm 0.4)$\%&$( 1.2\pm 0.2)$\%\\
\nodata&      $0.01$     &$(20.8\pm 0.6)$\%&$( 4.6\pm 0.3)$\%&$( 0.4\pm 0.1)$\%\\
\nodata&      $0.02$     &$(13.7\pm 0.5)$\%&$( 2.4\pm 0.2)$\%&$( 0.2\pm 0.1)$\%
\enddata
\tablenotetext{a}{200 uniform sampling over the period of [$t_0-t_{\rm E}$,
$t_0 + 3 t_{\rm E}$]; sampling interval of $0.02 t_{\rm E}$.}
\tablenotetext{b}{200 uniform sampling over the period of [$t_0-0.5 t_{\rm E}$,
$t_0 + 1.5 t_{\rm E}$]; sampling interval of $0.01 t_{\rm E}$.}
\tablecomments{
The definition of a detection is given in equation~(\ref{eq15}).
Here the degrees of the freedom of the fitting is 195: 200 data points and 
five-parameter photometric fitting. Our Monte-Carlo simulations of single 
lens events produce eight false detections out of 5,000 events so that the 
detection criterion corresponds to the rejection of the single lens model 
with 99.8\% confidence level. The uncertainty associated with the number 
of detections are determined by the Poisson statistics \citep{bevington92}, 
i.e.,
$\sqrt{N_{\rm detection}(N_{\rm total}-N_{\rm detection})/N_{\rm total}}$.
}\end{deluxetable}

\begin{deluxetable}{ccrrr}
\tablecaption{
Astrometric wide binary detection rate
\label{tab2}}
\tablehead{
\colhead{$d$}&
\colhead{astrometric uncertainty ($\theta_{{\rm E},1}$)}&
\colhead{$q=1.0$}&
\colhead{$q=0.316$}&
\colhead{$q=0.1$}}
\startdata
\cutinhead{`Broad' Sampling\tablenotemark{a}}
  $5.0$&$0.01$&$(100.\pm 0.0)$\%&$(100.\pm 0.0)$\%&$(67.4\pm 0.7)$\%\\
\nodata&$0.02$&$(100.\pm 0.0)$\%&$(91.7\pm 0.4)$\%&$(25.1\pm 0.6)$\%\\
\nodata&$0.03$&$(99.9\pm 0.1)$\%&$(69.4\pm 0.7)$\%&$(12.4\pm 0.5)$\%\\
  $8.0$&$0.01$&$(100.\pm 0.0)$\%&$(79.5\pm 0.6)$\%&$(17.5\pm 0.5)$\%\\
\nodata&$0.02$&$(96.1\pm 0.3)$\%&$(34.8\pm 0.7)$\%&$( 5.5\pm 0.3)$\%\\
\nodata&$0.03$&$(80.1\pm 0.6)$\%&$(18.3\pm 0.5)$\%&$( 2.2\pm 0.2)$\%\\
 $11.0$&$0.01$&$(98.3\pm 0.2)$\%&$(38.3\pm 0.7)$\%&$( 5.9\pm 0.3)$\%\\
\nodata&$0.02$&$(67.5\pm 0.7)$\%&$(11.7\pm 0.5)$\%&$( 1.6\pm 0.2)$\%\\
\nodata&$0.03$&$(41.4\pm 0.7)$\%&$( 5.8\pm 0.3)$\%&$( 0.7\pm 0.1)$\%\\
\cutinhead{`Dense' Sampling\tablenotemark{b}}
  $5.0$&$0.01$&$(100.\pm 0.0)$\%&$(98.3\pm 0.2)$\%&$(67.7\pm 0.7)$\%\\
\nodata&$0.02$&$(99.2\pm 0.1)$\%&$(83.2\pm 0.5)$\%&$(38.0\pm 0.7)$\%\\
\nodata&$0.03$&$(96.9\pm 0.2)$\%&$(68.0\pm 0.7)$\%&$(22.9\pm 0.6)$\%\\
  $8.0$&$0.01$&$(98.7\pm 0.2)$\%&$(77.2\pm 0.6)$\%&$(28.5\pm 0.6)$\%\\
\nodata&$0.02$&$(86.9\pm 0.5)$\%&$(48.2\pm 0.7)$\%&$( 8.0\pm 0.4)$\%\\
\nodata&$0.03$&$(74.8\pm 0.6)$\%&$(29.8\pm 0.6)$\%&$( 4.9\pm 0.3)$\%\\
 $11.0$&$0.01$&$(88.9\pm 0.4)$\%&$(51.2\pm 0.7)$\%&$( 9.8\pm 0.4)$\%\\
\nodata&$0.02$&$(66.3\pm 0.7)$\%&$(20.0\pm 0.6)$\%&$( 3.1\pm 0.2)$\%\\
\nodata&$0.03$&$(50.6\pm 0.7)$\%&$(10.2\pm 0.4)$\%&$( 1.4\pm 0.2)$\% 
\enddata
\tablenotetext{{\rm a,b}}{Same as in Table~\ref{tab1}.}
\tablecomments{
The detection is defined in the same way as in Table 1, but the degrees of
the freedom, here, is 393: 200 data points with two degrees of the freedom
and seven-parameter astrometric fitting.  The number of false detections
was six out of 5,000 simulated single lens events, which is marginally
different from the confidence level of the photometric detection.
}\end{deluxetable}

\begin{figure}
\plotone{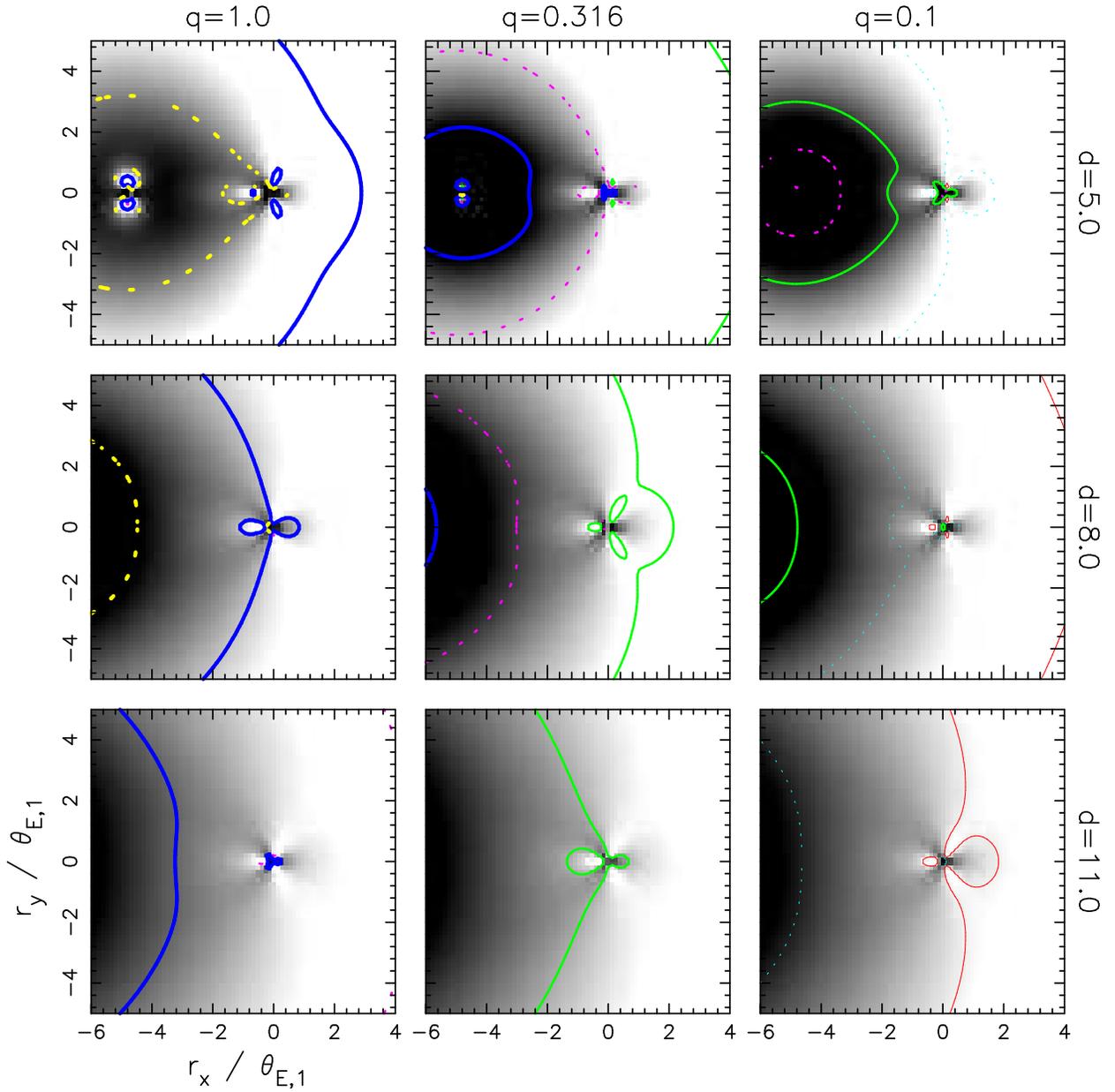}
\caption{
Contours of centroid shift deviations, $\Delta\delta$ (see eq.~[\ref{eq13}]),
for the same wide binaries as in Figs.~\ref{fig3} and \ref{fig4} but in 
the wider region around primary. The coordinate systems are the same as 
in Figs.~\ref{fig3} and \ref{fig4}.  Contours are drawn at the levels of
$\log(\Delta\delta/\theta_{{\rm E},1})= -2.1,\ -1.8,\ -1.5,\ -1.2,\ -0.9,
\ -0.6$ with colors of {\it Red, Cyan, Green, Magenta, Blue, Yellow}.  
Grey scale is used to provide the idea of the gradient over the region, 
but does not necessarily correspond to the absolute values of deviations.
\label{fig5}}\end{figure}

\end{document}